\documentclass[11pt,onecolumn, doublespacing]{IEEEtran}
\usepackage{textcomp}
\usepackage{xcolor}
\def\BibTeX{{\rm B\kern-.05em{\sc i\kern-.025em b}\kern-.08em
    T\kern-.1667em\lower.7ex\hbox{E}\kern-.125emX}}

\usepackage{amssymb,amsthm,amsmath,amsfonts}
\usepackage{enumerate, enumitem}
\usepackage{textcomp}
\usepackage{xcolor}
\usepackage{graphicx}

\usepackage{algorithm}
\usepackage{float}
\usepackage{epstopdf, epsfig}
\usepackage{tcolorbox}
\usepackage{setspace}
\usepackage[noend]{algpseudocode}
\usepackage{cite}
\usepackage{soul}

\newcommand{\HL}{\textcolor{black}}

\newtheorem{theorem}{Theorem}
\newtheorem{claim}{Claim}

\usepackage{hyperref}
\usepackage[justification=centering]{caption}

\graphicspath{{Figures/}}

\def\rbf{{\textbf r}}
\def\xbf{{\textbf x}}

\def\Xbf{{\textbf X}}

\newcommand{\orcid}[1]{\href{https://orcid.org/#1}  { {\includegraphics[scale=0.5]{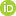}}}}

\begin{document}

\title{Joint Scheduling and Resource Allocation for Packets with Deadlines and Priorities}
\author{
  Li-on Raviv  \orcid{0000-0001-7526-7357}, \textit{ Member, IEEE,} \and
  Amir Leshem \orcid{0000-0002-2265-7463}, \textit{ Fellow, IEEE}
\thanks{Li-on Raviv is with the Faculty of Engineering, Bar-Ilan University, Ramat Gan 52900, Israel. Email: li.raviv@hotmail.com,} \\
\thanks{Amir Leshem is with the Faculty of Engineering, Bar-Ilan University, Ramat Gan 52900, Israel. Email: amir.leshem@biu.ac.il}}

\maketitle

\begin{abstract}
\HL{
Cellular networks provide communication for different applications. Some applications have strict and very short latency requirements, while others require high bandwidth with varying priorities. The challenge of satisfying the requirements grows in congested traffic where some packets might miss their deadlines.
Unfortunately, we prove that the problem is NP-Hard. To overcome this, we propose a new scheduling policy for packets with multiple priorities, latency requirements, and strict deadlines. To alleviate the complexity, our solution incorporates a novel time domain relaxation solved by linear programming.
Simulation results show that this method outperforms existing scheduling strategies.
}
\end{abstract}


\section{Introduction}
\label{sec:intro}
\HL{
Future and $5^{th}$ generation cellular networks aim to provide communication for people, machines, and devices, known as the Internet-of-Things.
New applications such as factory automation and intelligent transportation systems dictate a new level of end-to-end Quality Of Service (QoS) \cite{parvez2018survey, she2021tutorial} resulting in ambitious bandwidth and delay constraints.
}
These challenges are currently solved using three classes of traffic: enhanced Mobile Broadband (eMBB), Massive Machine Type Communications (mMTC), and Ultra-Reliable Low-Latency Communications (URLLC). The eMBB requires very high data rates with moderate latency. The mMTC requires low bandwidth, high connection density, enhanced coverage, and low energy consumption at the user end.
The URLLC requires extremely low delays with very high-reliability \cite{popovski2018wireless}. 
\HL{
Unfortunately, eMBB and mMTC packets scheduling does not support strict deadlines. We propose to adopt a new class of packets, extending the eMBB and the mMTC packets by adding deadlines. In addition, priorities are added to the packets allowing service reliability differentiation in times of congestion.
In addition, priorities are added to the packets allowing service reliability differentiation in times of congestion.
}

\subsection{State-of-the-art Scheduling Policies}
\label{sec:Policies}

In this section, we present state-of-the-art policies. We begin with non-preemptive policies for non-URLLC and continue with preemptive policies for URLLC.
The optimal policy Earliest Deadline First (EDF) ~\cite{stankovic2012deadline, capozzi2013downlink}, is commonly used in real-time. 
The EDF selects the packet with the earliest deadline for transmission.
Maximal Rate (MxRate) policy \cite{capozzi2013downlink} maximizes the overall transmission rate by assigning a Resource Block (RB) to a Mobile Subscriber (MS) that achieves the maximum transmission rate. 
The Modified Largest Weighted Delay First (M-LWDF) is a channel-aware scheduling policy that provides a bounded packet delivering delay \cite{capozzi2013downlink}. The scheduler assumes a deterministic deadline for each MS and an objective probability of meeting it. The M-LWDF assigns an RB to an MS according to its calculated metric. 
The Maximum Utility with Deadlines (MUD) policy \cite{raviv2018max}, 
\HL{transmits the packet with the maximal reward per unit time (reward rate). 
This approach is near optimal when queues are never empty}. 
Many schedulers supporting URLLC and eMBB have been proposed for 5G networks. 
The strict URLLC QoS objectives force the designer to optimize RB allocation while minimizing the loss of the eMBB throughput. 
In \cite{you2018resource} the proposed scheduler is based on utility values generated from linear programming (LP) relaxation and the Lagrangian dual.
In \cite{chang2019optimizing} the problem is formulated as a non-convex optimization problem for maximization of the weighted system throughput subject to QoS constraints.
\HL{
Lately, deep learning approaches for URLLC scheduling were introduced. These approaches are based on a deep deterministic policy gradient (DDPG) and its improvements \cite{li2020deep,gu2021knowledge}.
"Although deep learning algorithms have shown significant potential, the application of deep learning in URLLCs is
not straightforward" \cite{she2021tutorial}.
}

\subsection{Contribution}
\HL{
In this paper, we present a model supporting the coexistence of URLLC and non-URLLC packets having priorities and deadlines in cellular networks. The policy handles these packet classes differently. 
Non-URLLC is handled in a non-preemptive manner while maximizing the packets' Reward Rate (RR). URLLC is handled in a preemptive manner minimizing the damage they cause to the already being transmitted packets while meeting the stringent delay constraints. We provide a novel and computationally efficient policy for this case. 
}
%packets with deadlines and priorities.
We prove that the problem of non-preemptive scheduling under deadlines and QoS constraints is indeed NP-hard, and we incorporate more involved relaxations of the NP-hard problem to make the scheduling tractable. While \cite{raviv2018max} (MUD) provides an optimal solution for two priorities and deterministic service time classes, our scheduler outperforms MUD when the service time is non-deterministic (which is typical in the wireless case). 
In contrast to prior work, our solution handles both multi-user URLLC packets and non-URLLC packets, where the non-URLLC packets have QoS requirements, reflected by both deadlines and priorities. This extends the URLLC and eMBB resource allocation problem.
Similar to other resource allocation problems, it is presented as an ILP problem.
We then relax the problem into a linear program by allowing Time Division Multiplexing (TDM) at the sub-frame level.
%This allows better utilization of the RBs.
This work performs a parallel packet assignment differently than sequential greedy policies that the authors presented in \cite{raviv2018max, hadar2018scheduling}, and \cite{raviv2019scheduling}.
We use the following notation:
boldface capital letters denote matrices, boldface lowercase letters denote vectors, and standard lowercase letters denote scalars. The superscript $\textbf{A}^T$ denotes the transpose of the matrix. 

%-------------------------------------------------------------------
\section{System Model and Problem Formulation}
\label{sec:problem}
Consider a Base Transceiver Station (BTS) transmitting using $K$ channels to $M$ Mobile Subscribers (MSs). The traffic to the MSs contains URLLC packets and non-URLLC packets. 
\HL{
Each channel represents a Resource Block (RB) in the frequency domain.
%i.e., 12 sub-carriers. 
The RB is composed of one frame in the time domain. The frame itself is divided into sub-frames and mini-slot.
The RB transmission rate is frequency-dependent and time-dependent. Therefore, the MS reports the BTS with its current RBs transmission rates, alternatively, in TDD networks, the BS can estimate the downlink rate based on the uplink rate. Acquiring CSI is part of the physical layer of any coherent communication system, and therefore, does not result in additional overhead.
It is assumed that the rates are constant during packet transmission time. 
Each arriving packet has a designated MS, a deadline, a priority, a flag specifying whether it is a URLLC packet, and a payload.
}
The deadline is the delivery time of the packet. 
The model is hard-real-time; i.e., the packet's priority becomes a reward if the packet is delivered on time. 

\begin{figure}
    \centering
  \includegraphics[width=0.55\columnwidth]{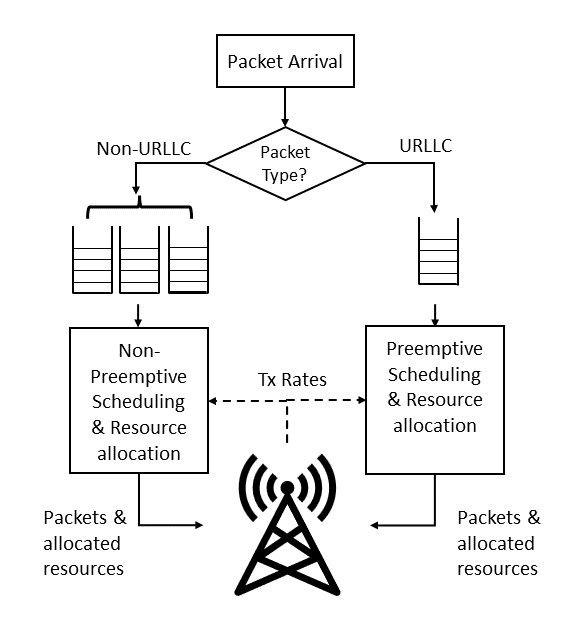}
  \caption{Packets Flow}
  \label{fig:URLLC}
\end{figure}

\HL{
The BTS comprises the queuing system, the scheduler, and the transmitter.
The BTS handles URLLC and non-URLLC traffic differently, as depicted in Fig. \ref{fig:URLLC}.
Packets that arrive at the BTS enter the queueing system. 
URLLC packets are stored in a single dedicated virtual queue, while the rest are stored in virtual queues according to their destination.} 
The URLLC virtual queue is ordered by First Come, First Served, while the scheduling policy orders the other queues.
\HL{
The scheduler implements a channel aware strategy choosing one or more packets based on their metadata for transmission.}
%The scheduler performs joint scheduling and resource allocation of one or more packets based on packet information and its RBs transmission rates.}
%\HL{\CL{Finally, the scheduler routes the packets to the transmitter with their channels and time allocations.}}
Non-URLLC packets are served in a non-preemptive manner and do not degrade the service to other packets.
By contrast, URLLC packets are served preemptively and can cause service degradation to other packets due to their strict latency and reliability requirements.
The server is never idle when there are packets in the queueing system (work conserving). 

The arrival process is modeled as follows: Let $J_i$ be the $i$-th packet that reached the BTS. $J_i$ is described by a tuple $<  a_i,e_i,m_i, l_i, w_i, u_i, \rbf_i > $
where $a_i\in \mathbb{R}^+$ is the arrival time and $e_i \in \mathbb{N}$ is the absolute expiration time in subframes resolution, $m_i \in\{ 1,\dots, M\}$ is the
designated MS, $l_i, w_i \in \mathbb{N}$ are the packet's length and reward, 
\HL{
$u_i$ is a boolean variable specifing if it is a URLLC packet ($u_i=T$).
$\rbf_i \in \mathbb{R}^{K\times 1}$ presents the transmission rates at the different channels at the current time.
}
\HL{
We measure policies' performance of non-URLLC and URLLC differently. URLLC allows one packet loss out of $10^5$ packets and any higher loss is a failure of the scheduler. The utility function for non-URLLC at time $t$ is:
\begin{equation}\label{eq:utility}
U(t) = \frac{1}{w(t)}\sum\limits_{J_i \in S_t}w_i.
\end{equation}}
\HL{
where $S_t$ is the set of non-URLLC packets that were delivered on time, until time $t$, and $w(t)=\sum_{a_i \leq t, ~u_i=F }w_i$.
Note that different reward schemes support a range of metrics.
For example, a determiistic reward $w_i=1$ measures the number of packets and setting $w_i=l_i$ measures the throughput.
}

\section{Maximal Reward Rate Policies}
\label{sec:Maximal}
\HL{
In this section, we propose a greedy approach that maximizes the rate of accumulated rewards for non-URLLC traffic. 
We assume that in highly congested networks, maximizing the reward rate (RR) accumulation leads to a maximal utility.
A similar approach is already used in greedy algorithms like MxRate.
Allocating channels to packets defines the packet's transmission rate and total reward rate, where the total reward rate is the sum of all channels' reward rates according to the channel allocation. 
The scheduler's objective is to maximize the total reward rate.
Let $t$ be the current time, and let $r_{i}^{min}=\frac{l_i}{e_i-t}$. Then, $r_{i}^{min}$ is the minimal expected transmission rate that guarantees timely delivery of packet $J_i$.
Non-preemptive schedulers do not degrade the transmission rate. Thus, if the allocated transmission rate is equal to or higher than the minimum transmission rate, the packet is delivered on time.
}
%$\sum_{i,k}\frac{w_i}{l_i}\rbf_i(k)$.
%, i.e., $\sum_{i,k}\frac{w_i}{l_i}\rbf_i(k)$
%The schedulers select packets and allocate them RBs, determining packets' transmission rates. 
%Let $t$ be the current time, then the minimal transmission rate required to deliver $J_i$ on time is $r_{i}^{min}=\frac{l_i}{e_i-t}$. 

\subsection{RB Allocation - Problem Formulation}

Let $\mathcal{J}$ be a set of eligible packets for transmission, i.e., packets that are at the head of the virtual queues. 
Our goal is to find a set $\hat{\mathcal{J}} \subseteq \mathcal{J}$ of packets and its RBs allocation that has the Maximal Reward Rate (MRR).
Let $\xbf_i \in \{0,1\}^{K \times 1}$ be a vector of optimization variables for packet $J_i$. RB $k$ is allocated to packet $J_i$ if $\xbf_i(k)=1$ and is not allocated if $\xbf_i(k)=0$. 
The problem can be formulated as follows:

\begin{subequations}
\begin{equation}\label{eq:subset}
\begin{aligned}
< \hat{\mathcal{J}}, \xbf_1 \dots \xbf_{|\hat{\mathcal{J}}|} >& = \underset{
\hat{\mathcal{J}} \subseteq \mathcal{J}, \xbf_1 \dots \xbf_{|\hat{\mathcal{J}}|}}{\arg\max} 
\{\sum\limits_{J_i \in \hat{\mathcal{J}}}\frac{w_i}{l_i} (\xbf^T_i \rbf_i)\}.
\end{aligned}
\end{equation}
\begin{equation}\label{eq:C1}
\begin{aligned}
\xbf_i^T \rbf_i & \geq \rbf^{min}_i,~i=1 \dots |\hat{\mathcal{J}}|.
\end{aligned}
\end{equation}
\begin{equation}\label{eq:C2}
\begin{aligned}
\sum\limits_{i=1}^{|\hat{\mathcal{J}}|} \xbf_i(k) \leq 1,~ k=1 \dots K.
\end{aligned}
\end{equation}
\end{subequations}
The objective function \eqref{eq:subset} maximizes the total reward rate, \eqref{eq:C1} are the deadline constraints, and \eqref{eq:C2} limits a RB allocation to a single packet.
The problem can be solved using ILP for each of the subsets of $\mathcal{J}$ by selecting the subset that has the total reward rate. The ILP formulation uses \eqref{eq:ILPMax} as the objective function, while preserving the linear constraints \eqref{eq:C1} and \eqref{eq:C2} for each subset $\hat{\mathcal{J}} \subseteq \mathcal{J}$.
\begin{equation}\label{eq:ILPMax}
\begin{aligned}
< \xbf_1 \dots \xbf_{|\hat{\mathcal{J}}|} >& = \underset{ \xbf_1 \dots \xbf_{|\hat{\mathcal{J}}|}}{\arg\max} 
\{\sum\limits_{J_i \in \hat{\mathcal{J}}}\frac{w_i}{l_i} (\xbf^T_i \rbf_i)\}.
\end{aligned}
\end{equation}
\HL{
Next, we show that the computational complexity of solving the above problem is NP-Hard.
}
\begin{theorem}
\label{claim:1}
Let $\mathcal{J}$ be a set of eligible packets, $\mathcal{K}$ be a set of available RBs, $\alpha \in \mathbb{R^+}$ and, let 
\begin{align}
    X_{\alpha}=\left\{ \bf{x}:  \sum\limits_{J_i \in \hat{\mathcal{J}}}\frac{w_i}{l_i} (\xbf^T_i \rbf_i) \ge \alpha \hbox{\ and\  (\ref{eq:C1})-(\ref{eq:C2}) hold} \right\}
\end{align}
then determining whether $X_{\alpha}=\emptyset$
is NP-Complete.
\end{theorem}
Proof: Assume that there is a solution to the problem. The computational complexity to verify that the solution is valid has polynomial time. Thus, the problem belongs to the complexity class NP.
We complete the proof by proving that the integer partition problem, which is known to be NP-Complete \cite{korf1998complete} can be reduced to this problem.
The integer partition problem assumes a set of $K$ integers $\{d_1,\dots d_K\}$. The problem is to determine whether there is a two-set partitioning such that the sum of the integers in both partitions is $\frac{1}{2}\sum_{i=1}^{K}d_i$ (the total sum is considered to be even).

The reduction is as follows:
Assume that there are two packets $J_1 \text{ and } J_2$ belonging to two different MSs.
The transmission rates of the RBs are $\rbf_i=(d_1,\dots, d_K)^T$, $r^{min}_i=\frac{1}{2}\sum_{j=1}^{K}d_j$ and $w_i=l_i=1$ for both packets ($i=1,~2$).
The objective variables $\xbf_1$ represents the first partitioning, i.e, if $x_1(k)=1$ then the $k$-th number is a member of the first partition and similarily $\xbf_2$ represents the second partition. Hence the partition problem is equivalent to solving the following problem:
\begin{subequations}
\begin{equation}\label{eq:RedPMax}
\begin{aligned}
< \xbf_1, \xbf_2 >& = \underset{ \xbf_1, \xbf_2}{\arg\max} 
\{\xbf^T_1 \rbf_1+\xbf^T_2 \rbf_2\}.
\end{aligned}
\end{equation}
\begin{equation}\label{eq:RedC1}
\begin{aligned}
\xbf_i^T \rbf_i & \geq \frac{1}{2}\sum\limits_{j=1}^{K}d_j,~i=1,~2.
\end{aligned}
\end{equation}
\begin{equation}\label{eq:RedC2}
\begin{aligned}
\xbf_1(k)+\xbf_2(k) \leq 1,~ k=1,\dots ,K.
\end{aligned}
\end{equation}
\end{subequations}
% If the objective function \eqref{eq:RedPMax} is equal to the sum of all integers there is a feasible partitioning into the set of integers. The objective function \eqref{eq:RedPMax} cannot be larger than the sum of all integers due to constraints \eqref{eq:RedC2}. Constraints \eqref{eq:RedC1} are equalities since $\sum_{j=1}^2r_i^{min}=\sum_{j=1}^{K}d_j$.
% Therefore solving the feasibility problem defined above provides a solution to the set partitioning problem, which implies that the computational complexity of $Dec(\alpha)$ is NP-Complete.
\begin{theorem}
The computational complexity of maximizing the total reward rate is NP-Hard.
\end{theorem}
Proof:
The computational complexity of finding the maximal total reward rate for a given set of packets is NP-Hard due to Claim \ref{claim:1}. As a result, the computational complexity to find the subset $\hat{\mathcal{J}} \subseteq \mathcal{J}$ with the maximal total reward rate is also NP-Hard.
$\square$
\subsection{ILP and its Polynomial Relaxation}
The number of times an ILP solver runs to find the Maximal total Reward Rate (MRR), given a set of eligible packets $\mathcal{J}$ and a set of available channels $\mathcal{K}$, is bounded by $O(2^{\min\{|\mathcal{J}|,|\mathcal{K}|\}})$.

The computational complexity can be reduced by observing that whenever $\hat{\mathcal{J}}$ has no feasible solution, also every  $\mathcal{I}$ such that $~\mathcal{I} \supset \hat{\mathcal{J}}$ has no feasible solution. Thus, computational complexity is reduced using a pruning process on the partial order of subsets under inclusion: if subsets are processed in ascending order of their size, and sets that contain a subset without a feasible solution are omitted.
Another method is to bound the size of the set; i.e., $|\hat{\mathcal{J}}| \leq p$ where $p < \min\{ |\mathcal{J}|, K\} $. This polynomial relaxation is marked as ILP($p$). The MUD, for example, is an ILP(1) relaxation which is a greedy policy.

\subsection{ILP Time Relaxation and LP Relaxation}
LP solvers have polynomial computational complexity \cite{gonzaga1995complexity}. Thus, using the LP relaxation ($\xbf_i \in [0,1]^{K \times 1}$) of Equation \eqref{eq:ILPMax} significantly reduces the computational complexity.
The results of the optimization variables can be fractions. These fractions define the time domain multiplexing. Each result states the fraction of time out of a one-time unit (e.g., a frame) in this approach. Alternatively, it defines the number of subframes allocated to the MS of each frame. Note that the standard relaxation using a threshold fails to deliver one of the packets on time.
\begin{claim}\label{claim:2subframes}
If the LP(2) solver assigns two packets to the same RB and the packets are transmitted one after the other in an EDF order, the first packet arrives on time, and two subframes bound the second packet's delay.
\end{claim}
Proof: Let $J_i:~i \in \{1,2\}$, be two packets that are allocated to the same RB. Let $d_i$ be the time left before the expiration time in subframes ($d_1 \leq d_2$). Let $x$ be the RB allocation for $J_1$ and let $\bar{x}=1-x$ be the RB allocation to $J_2$. 
Then $J_1, ~\lceil x d_1\rceil \leq  d_1$. 
$J_2$ ends its transmission at $\lceil \bar{x}d_2\rceil + \lceil x d_1\rceil \leq \bar{x} d_2 + x d_1 +2 \leq (\bar{x}+x) \max\{d_1,d_2\}+2= d_2+2$. $\square$

Thus, to meet the deadline, we need to deduct two subframes from the deadline of the packet with the latest expiry time.
As a result, the minimum rate becomes $\rbf_i^{min}=\frac{l_i}{\max\{e_i-t-2, 1\}}$.
The LP time relaxation conceals an inefficacy during the period between $d_1$ and $d_2$. In this period, the transmitter only uses the $\bar{x}$ fraction of the RB instead of the full capacity. Allocating the complete RB to the transmission of $J_2$ adds the capacity of $x (d_2-d_1)$ and if $x (d_2-d_1) \geq 2$ so that the deduction of two subframes from $d_2$ is redundant. Thus, the RB efficiency of transmitting two packets is $\frac{d_2}{d_2+2\Delta}$ where $\Delta$ is the length of a subframe.
The minimum rate can be adjusted to:
\begin{equation} %\label{eq:Seq2}
\begin{aligned}
 r_1^{min} = \begin{cases}
        \frac{l_1}{d_1}     & \text{where } x (d_2-d_1) \geq 2
        \\
        \frac{l_1}{d_1-2}   & \text{where } d_1 > 2 
        \\
        \infty              & \text{Otherwise}
        \end{cases}
\end{aligned}
\end{equation}

Similar to ILP($p$) polynomial reduction, we propose a complexity reduction by running the LP solver only on subsets of size 2, i.e., LP(2).
\HL{
Let $\xbf_1 \in [0, 1]^{K\times1}$ be the time allocation to $J_1$. In the optimal solution $x_1(k)+x_2(k) =1$ since otherwise  resource $k$ is not fully utilized. Therefore, $x_2(k) =1 – x_1(k)$. In vector form this results in $\xbf_2=1-\xbf_1$. Eq \eqref{eq:ILPMax} now becomes:
\begin{align}\label{eq:LPmax}
\nonumber \xbf & = \underset{\xbf_1 }{\arg\max} \{
\frac{w_1}{l_1} (\xbf_1^T \rbf_1)+\frac{w_2}{l_2} ((\boldsymbol{1}-\xbf_1)^T \rbf_2)\}\\
 & = \underset{ \xbf_1 }{\arg\max} \{
\xbf^T(\frac{w_1}{l_1} \rbf_1 - \frac{w_2}{l_2}\rbf_2)\}.\\
\nonumber \hbox{subject to: \ \ }&  \xbf_1^T \rbf_1 \geq r^{min}_1,~
%\text{ and }
\nonumber -\xbf_1^T  \rbf_2  \geq r^{min}_2-{\bf 1}^T {\rbf}_2.
\end{align}
}
Alg. \ref{alg:LP} solves \eqref{eq:LPmax} finds the best subset of packets that provides the maximal total reward rate and is used by the MRR-LP(2) policy. The complexity of Alg. \ref{alg:LP} is composed of the number of times the LP solver runs and the computational complexity of a single LP solver run.
The number of eligible packets defines the number of runs of the LP solver and is bounded by $M$. The policy checks the transmission's reward rate of at most two packets. Hence, the number of times the LP solvers run can be bounded by $O(M^2)$.
The computational complexity of an LP solver is $O((m+n)^{1.5}nL)$
\cite{vaidya1989speeding}, where $n$ represents the number of variables, $m$ presents the number of constraints, and $L$ represents the order of space required to input the problem. 
$K$ bounds the number of variables. Thus, the computational complexity is $O(K^{2.5}L)$. The overall complexity is $O(M^2K^{2.5}L)$. 
Unlike schedulers that allocate the resources in the frequency and time domains, the MRR-LP(2) only uses the frequency domain.
\HL{This significantly reduces the number of optimization variables and eliminates a time limit caused by the number of variables.
}
%which in turn reduces the computational complexity.}
\section{5G New Radio and beyond: URLLC}
\label{subsec:URLLC}
The MRR-LP(2) policy is a non-preemptive policy designed to support traffic with moderate delay requirements. In our model, the URLLC traffic has a dedicated virtual queue allowing enforcement of a different policy.
This section presents a policy designed to meet the strict QoS requirements of URLLC traffic.
%The URLLC virtual queue is ordered as FCFS. This order is equivalent to EDF due to the deterministic deadlines of the URLLC traffic.
% As mentioned earlier, URLLC traffic is immediately scheduled upon arrival, in the next minislot, i.e, no queuing is allowed. Thus if demands exceed the system capacity on a given minislot traffic would be lost.
URLLC schedulers use instant scheduling \cite{ji2018ultra}, i.e., they transmit in the next minislot while the rest of the traffic can only use  a slot level transmission. Furthermore, the traffic is pre-emptively overlapped at the minislot timescale, resulting in selective superposition/puncturing of non-URLLC allocations \cite{anand2020joint}. 
During the transmission time, the non-URLLC traffic suffers from service degradation, which all schedulers attempt to minimize. In our case, the target is to maximize the reward rate under the constraints of the deadlines.
Let $J_1,\dots J_{j}$ be non-URLLC packets that are being transmitted. Let $J_{j+1},\dots J_n$ be the new URLLC packets. 
We assume that the BTS is designed to handle all arriving URLLC traffic.
Let $\xbf_i$ be the assignment vector of the $i$-th packet to the $K$ RBs. The objective function for Maximizing the reward rate after allocating the URLLC packets is: 

\begin{align}
\label{eq:ULLCobj1}
< \xbf_1, \dots \xbf_n >& = \underset{
 \xbf_1 \dots \xbf_j}{\arg\max} 
\sum\limits_{i=1}^j\frac{w_i}{l_i} (\xbf^T_i \rbf_i).
\end{align}
The constraints are:
\begin{subequations}
\begin{align}
\label{eq:ULLCC1}
\xbf_i^T (\rbf_i-\sum\limits_{l=j+1}^{n} \frac{d_l}{d_i}\xbf_l ) & \geq \rbf^{min}_i,~i \leq j \text{ and } |\xbf_i| > 0.\\
\label{eq:ULLCC2}
\xbf_i^T \rbf_i & \geq \rbf^{min}_i,~ j < i \leq n\\
\label{eq:ULLCC3}
\sum\limits_{i=1}^{j} \xbf_i(k) \leq 1,&\sum\limits_{i=j+1}^{n} \xbf_i(k) \leq 1,~  1 \leq k \leq K,
\end{align}
\end{subequations}
where, \eqref{eq:ULLCC1} and \eqref{eq:ULLCC2} guarantee that all packets meet their deadlines subject to the URLLC RB assignment and \eqref{eq:ULLCC3} specify that at most one packet out of $J_1, \dots J_{n-1}$ is allocated to a single RB.
The problem is solvable by complete enumeration in exponential time ($2^{K|\mathcal{J}|}$) since the optimization variables are $0/1$.
A more straightforward approach is a policy that minimizes the number of overlapping RBs. The policy sorts the RBs according to the transmission rates of the URLLC packets in descending order. Then, the policy assigns RBs to URLLC packets until the packets' minimal rate is achieved. The RBs are allocated time-wise to the packet's deadline. If the minimum transmission rate is achieved, all other assignments to this packet are ignored.
Each RB is allocated time-wise to the packet's deadline. If the minimum transmission rate of a packet is achieved, then all other assignments to this packet are ignored.
The computational complexity of this policy is  O($K(n-j)\log (K(n-j))$), where $j$ is the number of non-URLLC packets.
%bounded ($\xbf_i \in \{0,1\}^{K \times 1}$). }

\section{Numerical Results}
\label{sec:Simulation}
\begin{figure} %[ht]
\centering
  \includegraphics[width=0.75\linewidth]{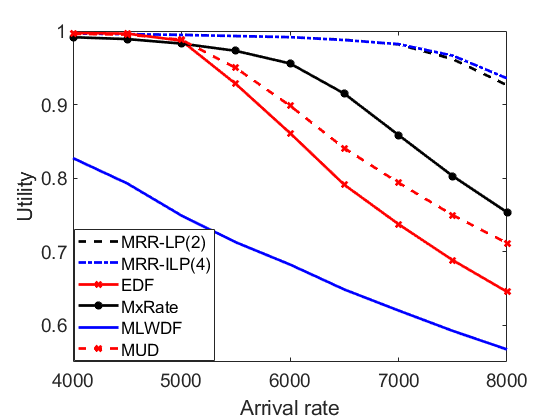}
  \centering
  \caption[center]{\small Utility Function at different arrival rates. }
  \label{fig:Set-1}
\end{figure}
In this section, we evaluate the performance of the proposed policies by dealing with both URLLC and non-URLLC traffic with priorities and deadlines. We simulated different types of packets and MSs transmission rate.
The transmission rate was simulated by tapped-delay channels, filtered by a multi-path Rayleigh fading channel. The parameters assumed an Urban Macro (city scenario) with no line of sight. The transmission tower height was 25m, and it transmitted at 42dBm. The 24 MSs were uniformly distributed in a range of 250m.  %similar to \cite{etsi138900}.
We used 15 RBs with a bandwidth of 180kHz at 6GHz and a 10ms frame, divided into ten subframes of 1ms.
We simulated a stream of packets composed of URLLC and five packet types with different traffic shares. Each packet type had different distributions of packet lengths, priorities, and deadlines as described in Table \ref{tbl:param}. 
The stream of packets arrival process was Poisson distributed with $\lambda_a$ parameter varying between 4,000 and 8,000 packets per second.
For each of the arrival rates, we performed 50 independent tests. Each test had 5,000 arriving packets.
We simulated six policies: MRR-LP(2), MRR-ILP(4), i.e., MRR policy with ILP(4) solver, MxRate, MUD, EDF, and M-LWDF, which were implemented similarly in the following~manner:
\begin{enumerate}[leftmargin=*,labelsep=5.8mm]
\item All policies used the URLLC preemptive policy \ref{subsec:URLLC}.
\item A packet whose deadline expired was dropped.
\end{enumerate}

%----------------------------------------------------------------
The results showed that all URLLC packets arrived on time for all scheduling schemes.
Figs. \ref{fig:Set-1} depicts the mean of the utility function \eqref{eq:utility}. 
We measured the percentage of delivered bytes at the different arrival rates, and the results were similar to the results of the utility function.
The performance of all policies decreases as the arrival rate increases.
Both MRR-LP(2) and MRR-ILP(4) utilities had similar performance.
These policies presented better performance than the alternatives. The difference between MRR-LP(2) and MRR-ILP(4) performance and other policies was high in heavy traffic. MRR-LP(2) and MRR-ILP(4) performance were $25\%$ better than other policies.
%All classical policies (EDF, MxRate, MUD, and MLWDF) performed significantly worse for arrival rates higher than 5,000 packets per second.
It can be seen that the proposed MRR-LP(2) policy significantly outperforms the classical policies (EDF, MxRate, MUD, and MLWDF) for arrival rates above 5,000 packets per second.
\HL{
We measured the distribution of the number of simultaneously added packets. 
MRR-ILP(4) added one or two packets in $95\%$ of the cases, three packets in $4\%$ of the cases, and four packets in the rest. The reward loss of MRR-ILP(2) was less than $1.5\%$ relative to MRR-ILP(4).
}
%--------------------------------------------------------------------

\begin{table}[htbp]
{\footnotesize
\captionsetup{font=footnotesize}
\caption{Traffic Parameters}
\begin{center}
\begin{tabular}{|c|c|c|c|c|}
\hline
%\textbf{Table}&\multicolumn{3}{|c|}{\textbf{Table Column Head}} \\
%\cline{2-4} 
\textbf{\textit{Packet}}&\textbf{\textit{Traffic}}& \textbf{\textit{Packet Len.}}  &  \textbf{\textit{Deadline}}& \textbf{\textit{Priority}}\\
\textbf{\textit{type}}&\textbf{\textit{share}}& \textbf{\textit{distrib. (Byte)}} & \textbf{\textit{ distrib. (sec)}}  &\textbf{\textit{per Byte }}\\
\hline
URLLC& $8\%$ & $32$ &  0.0005 &  $\infty$   \\ 
\hline
$1$& $13.8\%$ & $64$ &  exp($0.1$) &  $4$   \\ 
\hline
$2$& $32.2\%$ & $\mathcal{U}(64,100)$ &  exp($0.2$) &  $1$   \\ 
\hline
$3$ & $4.6\%$ & $\mathcal{U}(100,1400)$ &   exp($0.2$) &  $2$  \\ 
\hline
$4$ & $18.4\%$ & $\mathcal{U}(100,1400)$ &   exp($0.3$) &  $1$   \\ 
\hline
$5$ & $23\%$ & $1500$ &  exp($0.4$) &  $1$   \\ 
\hline
\end{tabular}
\label{tbl:param}
\end{center}
}
\end{table}

\begin{algorithm}[H]
\caption{MRR-LP(2)} \label{alg:LP}
\textbf{Input: } $\mathcal{J}$ is a set of eligible packets for transmission.\\
\textbf{Output:} $best\hat{\mathcal{J}}$ is the subset of $\mathcal{J}$ with the MRR. $bestX$ is the allocation of packets to subcarriers.

\begin{algorithmic}[1]
\Procedure{MRR-LP}{$\mathcal{J}$}
\For {$i \leftarrow 1:|\mathcal{J}|$}
    \State $\hat{\mathcal{J}}=\{J_i\}$,
    \State Allocate all available RBs to $J_i$.
    \If{there is a feasible solution}
        \If {$\hat{\mathcal{J}}$ has higher RR than $best\hat{\mathcal{J}}$}
            \State $best\hat{\mathcal{J}}=\hat{\mathcal{J}}$,~
             $bestX\leftarrow \boldsymbol{1}$
        \EndIf
        \For {$j\leftarrow i+1:|\mathcal{J}|$}
            \State $\hat{\mathcal{J}}=\{J_i,~J_j\}$
            % omitting RBs where $\rbf_i(k),\rbf_j(k)=0$.
            \State Compute $\Xbf$ by solving \eqref{eq:LPmax}. 
            \If{$\hat{\mathcal{J}}$ RR is higher than $best\hat{\mathcal{J}}$}
                \State $best\hat{\mathcal{J}}\leftarrow\hat{\mathcal{J}}$,~
                $bestX\leftarrow \Xbf$
            \EndIf
        \EndFor
    \EndIf
\EndFor
\State \textbf{return} $best\hat{\mathcal{J}},~bestX$
\EndProcedure
%\vspace{6pt}
\end{algorithmic}
\end{algorithm}

%\begin{figure}%[ht]
%\centering
%  \includegraphics[width=0.76\linewidth]{bytes 18-05-2022.png}
%  \centering
%  \caption[center]{\small Relative Delivered bytes at different arrival rates. }
%  \label{fig:Set-2}
%\end{figure}
\section{Conclusion}
\label{sec:Conclusions}
This paper presents a novel joint scheduling and resource allocation policy for URLLC and non-URLLC with priorities and deadlines.
The general problem is shown to be NP-hard.
We proposed a policy that performs a relaxation based on LP with time-division multiplexing. Simulations showed that the policy significantly outperforms the classical policies over the 5G channels and traffic model.

%\bibliographystyle{IEEEbib}
%\bibliography{Bib210713.bib}

\begin{thebibliography}{10}

\bibitem{parvez2018survey}
Imtiaz Parvez, Ali Rahmati, Ismail Guvenc, Arif~I Sarwat, and Huaiyu Dai,
\newblock ``A survey on low latency towards {5G: RAN}, core network and caching
  solutions,''
\newblock {\em IEEE Communications Surveys \& Tutorials}, vol. 20, no. 4, pp.
  3098--3130, 2018.

\bibitem{she2021tutorial}
Changyang She, Chengjian Sun, Zhouyou Gu, Yonghui Li, Chenyang Yang, H~Vincent
  Poor, and Branka Vucetic,
\newblock ``A tutorial on ultrareliable and low-latency communications in 6g:
  Integrating domain knowledge into deep learning,''
\newblock {\em Proceedings of the IEEE}, vol. 109, no. 3, pp. 204--246, 2021.

\bibitem{popovski2018wireless}
Petar Popovski, Jimmy~J Nielsen, Cedomir Stefanovic, Elisabeth De~Carvalho,
  Erik Strom, Kasper~F Trillingsgaard, Alexandru-Sabin Bana, Dong~Min Kim,
  Radoslaw Kotaba, Jihong Park, et~al.,
\newblock ``Wireless access for ultra-reliable low-latency communication:
  Principles and building blocks,''
\newblock {\em IEEE Network}, vol. 32, no. 2, pp. 16--23, 2018.

\bibitem{stankovic2012deadline}
John~A Stankovic, Marco Spuri, Krithi Ramamritham, and Giorgio~C Buttazzo,
\newblock {\em Deadline scheduling for real-time systems: {EDF} and related
  algorithms}, vol. 460,
\newblock Springer Science \& Business Media, 2012.

\bibitem{capozzi2013downlink}
Francesco Capozzi, Giuseppe Piro, Luigi~Alfredo Grieco, Gennaro Boggia, and
  Pietro Camarda,
\newblock ``Downlink packet scheduling in {LTE} cellular networks: Key design
  issues and a survey,''
\newblock {\em IEEE Communications Surveys \& Tutorials}, vol. 15, no. 2, pp.
  678--700, 2013.

\bibitem{raviv2018max}
{Li-on} Raviv and Amir Leshem,
\newblock ``{M}aximizing service reward for queues with deadlines,''
\newblock {\em IEEE/ACM Transactions on Networking}, vol. 26, no. 5, pp.
  2296--2308, Oct 2018.

\bibitem{you2018resource}
Lei You, Qi~Liao, Nikolaos Pappas, and Di~Yuan,
\newblock ``Resource optimization with flexible numerology and frame structure
  for heterogeneous services,''
\newblock {\em IEEE Communications Letters}, vol. 22, no. 12, pp. 2579--2582,
  2018.

\bibitem{chang2019optimizing}
Bo~Chang, Lei Zhang, Liying Li, Guodong Zhao, and Zhi Chen,
\newblock ``Optimizing resource allocation in {URLLC} for real-time wireless
  control systems,''
\newblock {\em IEEE Transactions on Vehicular Technology}, vol. 68, no. 9, pp.
  8916--8927, 2019.

\bibitem{li2020deep}
Jing Li and Xing Zhang,
\newblock ``Deep reinforcement learning-based joint scheduling of {eMBB and
  URLLC} in {5G} networks,''
\newblock {\em IEEE Wireless Communications Letters}, vol. 9, no. 9, pp.
  1543--1546, 2020.

\bibitem{gu2021knowledge}
Zhouyou Gu, Changyang She, Wibowo Hardjawana, Simon Lumb, David McKechnie, Todd
  Essery, and Branka Vucetic,
\newblock ``Knowledge-assisted deep reinforcement learning in 5g scheduler
  design: From theoretical framework to implementation,''
\newblock {\em IEEE Journal on Selected Areas in Communications}, vol. 39, no.
  7, pp. 2014--2028, 2021.

\bibitem{hadar2018scheduling}
Ido Hadar, {Li-on} Raviv, and Amir Leshem,
\newblock ``Scheduling for {5G} cellular networks with priority and deadline
  constraints,''
\newblock in {\em 2018 IEEE International Conference on the Science of
  Electrical Engineering in Israel (ICSEE)}. IEEE, 2018, pp. 1--5.

\bibitem{raviv2019scheduling}
Li-on Raviv and Amir Leshem,
\newblock ``Scheduling for multi-user multi-input multi-output wireless
  networks with priorities and deadlines,''
\newblock {\em Future Internet}, vol. 11, no. 8, pp. 172, 2019.

\bibitem{korf1998complete}
Richard~E Korf,
\newblock ``A complete anytime algorithm for number partitioning,''
\newblock {\em Artificial Intelligence}, vol. 106, no. 2, pp. 181--203, 1998.

\bibitem{gonzaga1995complexity}
Clovis~C Gonzaga,
\newblock ``On the complexity of linear programming,''
\newblock {\em Resenhas do Instituto de Matem{\'a}tica e Estat{\'\i}stica da
  Universidade de S{\~a}o Paulo}, vol. 2, no. 2, pp. 197--207, 1995.

\bibitem{vaidya1989speeding}
Pravin~M Vaidya,
\newblock ``Speeding-up linear programming using fast matrix multiplication,''
\newblock in {\em 30th annual symposium on foundations of computer science}.
  IEEE Computer Society, 1989, pp. 332--337.

\bibitem{ji2018ultra}
Hyoungju Ji, Sunho Park, Jeongho Yeo, Younsun Kim, Juho Lee, and Byonghyo Shim,
\newblock ``Ultra-reliable and low-latency communications in {5G} downlink:
  Physical layer aspects,''
\newblock {\em IEEE Wireless Communications}, vol. 25, no. 3, pp. 124--130,
  2018.

\bibitem{anand2020joint}
Arjun Anand, Gustavo De~Veciana, and Sanjay Shakkottai,
\newblock ``Joint scheduling of {URLLC and eMBB} traffic in {5G} wireless
  networks,''
\newblock {\em IEEE/ACM Transactions on Networking}, vol. 28, no. 2, pp.
  477--490, 2020.

\end{thebibliography}

\end{document}